# Fourier analysis of near-field patterns generated by propagating polaritons

Minsoo Jang[1,†], Sergey G. Menabde[1,†,*], Fatemeh Kiani[2], Jacob T. Heiden[1], Vladimir A. Zenin[3], N. Asger Mortensen[4,5], Giulia Tagliabue[2], and Min Seok Jang[1,*]

[1]School of Electrical Engineering, Korea Advanced Institute of Science and Technology (KAIST), Daejeon 34141, Korea

[2]Laboratory of Nanoscience for Energy Technologies (LNET), School of Engineering (STI), EPFL, 1015 Lausanne, Switzerland

[3]Center for Nano Optics, University of Southern Denmark, Odense, DK-5230, Denmark

[4]POLIMA – Center for Polariton-driven Light-Matter Interactions, University of Southern Denmark, DK-5230 Odense, Denmark

[5]Danish Institute for Advanced Study, University of Southern Denmark, DK-5230 Odense, Denmark

†: These authors contributed equally to this work

*: menabde@kaist.ac.kr (S. G. M.), and jang.minseok@kaist.ac.kr (M. S. J.)

**Abstract**

Scattering-type scanning near-field optical microscope (s-SNOM) has become an essential tool to study polaritons – quasiparticles of light coupled to collective charge oscillations – via direct probing of their near field with a spatial resolution far beyond the diffraction limit. However, extraction of the polariton's complex propagation constant from the near-field images requires subtle considerations that have not received necessary attention so far. In this study, we discuss important yet overlooked aspects of the near-field analysis. First, we experimentally demonstrate that the sample orientation inside the s-SNOM may significantly affect the near-field interference pattern of mid-infrared polaritons, leading to an error in momentum measurement up to 7.7% even for the modes with effective index of 12.5. Second, we establish a methodology to correctly extract the polariton damping rate from the interference fringes depending on their origin – the s-SNOM nano-tip or the material edge. Overall, our work provides a unified framework for the accurate extraction of the polariton momentum and damping from the near-field interference fringes.

# I. INTRODUCTION

Material characterization is crucial for advancing technologies in a wide range of fields; this holds true for optics as well. In particular, polaritons, the quasi-particles born out of coupling between light and excitations in material (collective oscillations of free carriers, phonons, excitons etc.), provide comprehensive information about the electrodynamic response of the employed material [1-9]. At the same time, low-dimensional surface polaritons promise diverse nanophotonic applications due to the strong field confinement, delivering the enhanced light-matter interaction [9-16]. The surface polaritons possess tremendous application potential such as high-density data storage [17], surface-enhanced Raman spectroscopy [18, 19], thermal emission control [20-22], negative refraction [23], biosensing [24, 25], and beam steering [26-28] .

Scattering-type scanning near-field optical microscope (s-SNOM) offers a breakthrough method for direct measurement of the polaritonic near-field amplitude and phase with a spatial resolution of inherently non-radiating excitations beyond the diffraction limit – a regime which cannot be probed by far-field instrumentation [29]. In a conventional s-SNOM experiment, both excitation and detection of surface polaritons are typically carried out via light scattering by the sharp nano-tip of the atomic force microscope (AFM), resulting in a near-field interference pattern which carries information about the complex propagation constant [1, 3, 30-37]. In this study, we rigorously analyze the extraction of the complex propagation constant from the near-field data, and demonstrate that commonly accepted approximations for the polariton momentum analysis may lead to significant quantitative errors.

While the conventional approach of direct fitting of the near-field signal can yield good qualitative results [33, 36, 38], we show that the Fourier analysis of the near-field interference in momentum space is a rigorous and convenient method readily applicable even for mixed signals formed by a superposition of several propagating modes. Importantly, we demonstrate that the polaritonic damping rate can be estimated from the near-field interference fringes as well, but is not given by the width of their Fourier spectrum as implicitly assumed in previous works [33, 39, 40]. We also highlight the different analysis methods for the polaritons launched by the s-SNOM nano-tip and material edge, required for the correct extraction of damping in each case. Furthermore, we elaborate on the usually dismissed effects of the azimuthal and incidence angles between the s-SNOM excitation beam and the sample when polaritons are launched by the material edge, and show that these effects should be considered for rigorous analysis of the near-field interference.

The importance of the sample orientation relative to excitation beam has been explored earlier for polaritons [8, 41-44] with relatively low effective indices, i.e. when $k_p(\omega) \sim k_0 = \omega/c$, where $k_p$ is the polariton momentum and $k_0$ is that in the free space. However, sample orientation has typically been neglected for near-field probing of the highly confined mid-infrared (mid-IR) polaritons, such as hyperbolic phonon-polaritons (HPP) in thin hexagonal boron nitride (hBN) [45] or alpha-phase molybdenum trioxide (α-MoO$_3$) [2, 37], including our own prior works on hBN [39] on α-MoO$_3$ [40], and many other works studying mid-IR polaritons.

For the experiments, we use hBN on evaporated and monocrystalline gold where HPP are excited by light scattering at the atomically clear edges of hBN or gold crystals [46]. We demonstrate the discrepancy of the measured polariton momentum up to 7.7% compared to the actual value, even when the HPP effective index $n_\text{eff}(\omega) = k_p(\omega)/k_0(\omega)$ is as large as 12.5, resulting merely from the change of the sample orientation. Our results are particularly important for a growing number of studies where the near-field data is used to measure properties of novel polaritonic materials [2, 4, 5, 37, 47, 48].

## II. NEAR-FIELD COMPONENTS OF POLARITON INTERFERENCE

The operation principle of s-SNOM is based on the interferometric detection of the near-field signal from a sample surface which is coupled to the far-field via scattering by a sharp AFM nano-tip while scanning the sample in a tapping mode [29]. The commonly used reflection mode of the s-SNOM setup exploits the same beam path for both illumination and collection of the scattering, therefore the excitation beam is constantly focused on the nano-tip, as illustrated in Fig. 1(a). However, in the mid-IR s-SNOM setup, the polaritonic material edge is also illuminated unless it is located at a considerable distance away from the tip. This is because the beam spot size is of the order of the free-space wavelength, $\lambda_0$, while the scan size (and, thus, the maximum distance between the edge and the tip) is of the order of the polariton wavelength, $\lambda_p$, which can be 10-100 times less than $\lambda_0$ due to the high effective index of the polariton mode. The beam scatters at both the metallic nano-tip and the material edge, resulting in a local electric field at the nano-tip, $\mathbf{E}_\text{tip}$, given as a superposition of the incident field, $\mathbf{E}_\text{i}$, and the polariton fields, $\mathbf{E}_\text{pt}$ and $\mathbf{E}_\text{pe}$, from the tip-launched and the edge-launched modes, respectively:

$$\mathbf{E}_\text{tip} = \mathbf{E}_\text{i} + \mathbf{E}_\text{pt} + \mathbf{E}_\text{pe}. \tag{1}$$

The resulting near-field interference pattern thus carries the information about the polariton mode, but generally consists of several components of different nature which must be disentangled to properly extract the complex propagation constant of the polariton.

The contribution of the tip-launched polariton is nonzero only when the excited polariton is returned back to the tip, for example, by reflecting from the material edge. By considering only a single straight material edge at a distance $x$ from the tip (Fig. 1(b)), the contribution of the tip-launched polariton can be expressed as:

$$\mathbf{E}_{\text{pt}}(x) = \tilde{c}_{\text{pt}}\tilde{r}\sqrt{\frac{\lambda_{\text{p}}}{2(x+\lambda_{\text{p}}/2)}}e^{i\tilde{k}_{\text{p}}2x}\mathbf{E}_{\text{i}}, \qquad (2)$$

where $\tilde{c}_{\text{pt}}$ is the complex excitation efficiency of the tip-launched polariton, $\tilde{r}$ is the complex polariton reflection coefficient at the edge, and $\tilde{k}_{\text{p}} = k_{\text{p}} + ik'_p$ is the complex polariton wavevector; $k_{\text{p}} = 2\pi/\lambda_{\text{p}}$. Since the tip-launched polariton has a radially diverging wavefront, an additional geometrical correction term $\sqrt{\lambda_{\text{p}}/(x+\lambda_{\text{p}}/2)}$ is included to account for the power conservation in the effectively 2D polaritonic wave that has initial mode size of $\sim\lambda_{\text{p}}$ at the moment of excitation ($x \geq \lambda_{\text{p}}/2$). We emphasize that this approach resolves the ambiguity of $x$-axis origin for the fringes analysis which is inevitable when a simple geometric correction parameter $1/\sqrt{x}$ is used [44].

At the same time, the edge-excited polariton exhibits different propagation dynamics in terms of the field damping and the wavefront shape. Considering the distance between the tip and the material edge, $x$, the incident field at the material edge is $\mathbf{E}_0 = \mathbf{E}_{\text{i}}\exp(-ik_0 x \cos\varphi \cos\alpha)$, where $\alpha$ is the angle of incidence to the material surface, and $\varphi$ is the azimuthal angle of incidence compared to the straight edge (Figs. 1(a)-1(c)). The phase delay $k_0 x \cos\varphi \cos\alpha$ is observed as a frequency shift in the momentum space. At the tip position, the field from the edge-launched polariton is given by [49]:

$$\mathbf{E}_{\text{pe}}(x) = \tilde{c}_{\text{pe}}e^{i\tilde{k}_{\text{p}}x}\mathbf{E}_0 = \tilde{c}_{\text{pe}}e^{i(\tilde{k}_{\text{p}}-\Delta k)x}\mathbf{E}_{\text{i}}, \qquad (3)$$

where $\tilde{c}_{\text{pe}}$ is the complex excitation efficiency of the edge-launched polariton, and $\Delta k = k_0 \cos\varphi \cos\alpha$ indicates the different phase of the excitation wavefront at the edge and tip positions that causes a shift of the spatial frequency of the interference pattern imaged by the s-SNOM. It should be noted that the reverse light pass, i.e., when the incident beam excites polaritons at the tip, which are then scattered by the edge to the far-field, is described with similar equation, therefore this contribution is incorporated into the arbitrary parameter $\tilde{c}_{\text{pe}}$.

The scattered field from the tip, $\mathbf{E}$, which is measured by the s-SNOM, is proportional to the local electric field $\mathbf{E}_{\text{tip}}$ under the tip (Eq. (1)) and its effective polarizability $\tilde{\alpha}_{\text{eff}}$ that describes the near-field coupling between the tip and the sample surface [49]. From Eq. (1)-(3), by recognizing that the tip-coupling and the edge-coupling of the polaritons are both highly inefficient processes (i.e., $|\tilde{c}_{\text{pt}}|$, $|\tilde{c}_{pe}| \ll 1$) [31] and taking the Taylor series expansion of $|\mathbf{E}|$ up to the first order terms in $g_0$ and $f_0$, $|\mathbf{E}|$ can be approximated as:

$$|\mathbf{E}| \approx |\tilde{\alpha}_{\text{eff}}| \left[ 1 + |\tilde{c}_{pe}| e^{-k'_p x} \cos[(k_p - \Delta k)x + \psi_{pe}] + |\tilde{c}_{pt}||\tilde{r}| \sqrt{\frac{\lambda_p}{2(x + \lambda_p/2)}} e^{-2k'_p x} \cos(2k_p + \psi_{pt}) \right] |\mathbf{E}_i|, \quad (4)$$

where $\psi_{pe} = \arg(\tilde{c}_{pe})$ and $\psi_{pt} = \arg(\tilde{c}_{pt}\tilde{r})$. The first term in Eq. (4) is the direct contribution of the incident beam, and the second and third oscillatory terms arise from the incident light interfering with the tip-launched polaritons and the edge-launched polaritons, respectively.

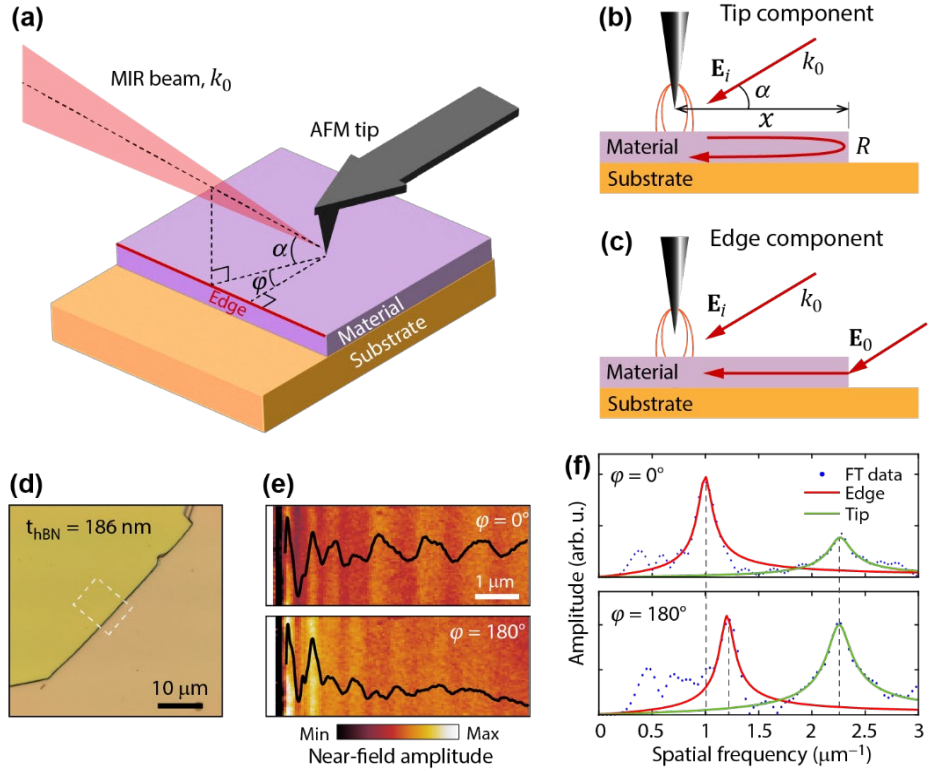

Fig. 1. Schematic model of s-SNOM and analyzing method. (a) Illustration of the experimental setup where p-polarized MIR beam with wavevector $k_0$ is focused on the AFM tip at the angle $\alpha = 30°$ with respect to material surface. The edge rotation angle $\varphi$ is defined as the angle between the plane of incidence and the normal to the edge. (b) and (c) are side view of the experimental setup showing the two excitation mechanisms: (b) by the AFM tip and (c) by the edge. (d) Optical microscope image of a 186 nm-thick hBN on evaporated gold substrate. (e) The near-field amplitude map at excitation wavenumber of 1500 cm$^{-1}$ and angle $\varphi = 0°$ and 180°. The black curve shows the averaged data in the direction perpendicular to the fringes. The s-SNOM images were obtained within the area marked by white dashed rectangle in (d). (f) Fourier spectra (blue dots) of the near-field amplitude in (e) integrated across the fringes. Lorentzian-like line shape is used to fit the spectra of the edge component (red) and the tip component (green). The black dashed line indicates the Lorentzian peak position.

## III. POLARITON MOMENTUM ANALYSIS

Fourier transformation (FT) provides a convenient method to analyze surface polariton properties from the measured near-field images. Most notably, it enables discrimination between the edge-launched and the tip-launched polariton contributions as they produce interference patterns with largely different momenta, $k_\text{p} - \Delta k$ and $2k_\text{p}$, respectively, as expected from Eq. (4). By separately analyzing the Fourier spectrum of each excitation component, it is possible to reliably and quantitatively determine the complex wavevector.

Based on the Eq. (4), $|\mathbf{E}|$ profile from the edge-launched component has the form of a damped harmonic oscillator, $f(x) = A\mathrm{e}^{-2\pi\gamma x}\cos(2\pi\nu x)$, with the analytically derived Fourier spectrum, $F(s)$, given by a Lorentzian-like function:

$$F(s) = A\left(\frac{\gamma + is}{(\gamma + is)^2 + \nu^2}\right). \tag{5}$$

Consequently, fitting the Fourier spectrum of the near-field interference pattern by Eq. (5) allows to extract not only the spatial frequency $\nu_\text{p} = k_\text{p}/2\pi$, but also the damping rate of the polariton $\gamma_\text{p} = k'_p/2\pi$. In the following analysis, we use Eq. (5) to fit Fourier spectra of the near-field pattern from both components. Although $f(x)$ is only an approximation for the signal from the edge-launched component, Eq. (5) still fits the Fourier spectrum well, accurately providing value of $\nu_\text{p}$.

To demonstrate the near-field signal from of the differently launched polaritons, we performed s-SNOM measurements near the edge of a 186 nm-thick hBN crystal on an evaporated gold substrate (Fig. 1(d)). When hBN is placed on gold, the HPP couples with its mirror image in the metal, producing an approximately twice more confined hyperbolic image phonon-polariton (HIP) which becomes the fundamental mode [50, 51].

Figure 1(e) shows the near-field amplitude maps in the vicinity of the hBN edge at $\alpha = 30°$ for $\varphi = 0°$ and $180°$ where tip- and edge-launched polaritons produce complex interference pattern. The analyzed interference profiles (black curves) are obtained by averaging the near-field maps in the direction parallel to the fringes. Baseline fitting by a low-order polynomial is employed to eliminate minor but inevitable background fluctuations, and then the Fourier spectrum of the near-field profile is calculated starting from the second maximum. The first maximum (closest to the edge) is mostly due to the strong near-field scattering at the sharp material edge and thus must be excluded from analysis.

In agreement with expectations based on Eq. (1), Fourier spectrum of the near-field amplitude (blue dotted lines in Fig. 1(f) and fitted Lorentzian-like line shapes) exhibits two clear peaks: the higher frequency peak centered at $2\nu_p = 2.26$ μm$^{-1}$ is due to the tip-launched polaritons and does not depend on $\varphi$, while the lower frequency peak due to the edge-launched polaritons is centered at $\nu_p \mp \Delta\nu = 0.99$ μm$^{-1}$ and 1.24 μm$^{-1}$ for $\varphi = 0°$ and 180°, respectively, showing a strong dependency on $\varphi$. The spatial frequency difference between the two angles is 0.25 μm$^{-1}$, which is very close to the expected value of $2\Delta\nu = 2\nu_0 \cos\alpha \cos\varphi = 0.26$ μm$^{-1}$, where $\nu_0 = k_0/2\pi = \lambda_0^{-1}$. Consequently, the spatial frequency of the HIP can be estimated from the edge-launched contribution when the angle-dependent frequency shift is accounted for, providing $\nu_p = 1.12$ μm$^{-1}$ which agrees well with $\nu_p = 1.13$ μm$^{-1}$ extracted from the tip-launched component.

The full width at half maximum (FWHM) of the fitted spectra of the edge-launched HIP (red Lorentzian-like line shapes in Fig. 1(f)) is almost the same for both $\varphi = 0°$ and 180°: FWHM = 0.22 μm$^{-1}$ and 0.24 μm$^{-1}$, respectively. The tip-launched HIP spectra (green Lorentzian-like profiles in Fig. 1(f)) are expected to be more than two times wider compared to those of the edge-launched HIP because of doubled damping $\gamma_p$ and the additional geometrical damping factor. Surprisingly, their FWHM are less than expected: 0.35 μm$^{-1}$ and 0.36 μm$^{-1}$ for $\varphi = 0°$ and 180°, respectively. We believe that this might be due to the dependence of $\tilde{c}_{pt}$, $\tilde{\alpha}_{eff}$, and $\tilde{r}$ on the tip-edge distance $x$, which is expected near the edge [52], where the fast-decaying tip-launched polaritons are mostly pronounced. Therefore, only the edge-launched HIP spectra can be used to accurately calculate the polariton damping, as will be discussed in details further in the text.

To experimentally estimate the range of the momentum extraction error as a function of angle $\varphi$, we use a very thin (≈20 nm) monocrystalline gold flake covered with hBN, as shown in Figs. 2(a)-2(b). We perform s-SNOM scans across the atomically clear edge of the gold crystal, which excites polariton modes with much higher efficiency than the AFM tip or the hBN edge, owing to its much larger scattering cross-section [31]. Furthermore, the tip-launched polaritons experience little reflection by the underlying gold edge (i.e., $|\tilde{r}| \ll 1$) and thus practically do not contribute to the interference pattern [39, 40]. Consequently, the near-field patterns arise mainly due to the interference between the edge-launched polaritons and the incident light, further simplifying the analysis.

Figure 2(c) displays the near-field maps over the 76 nm-thick hBN flake at the excitation frequency of 1530 cm$^{-1}$ for various angles $\varphi$. The Fourier spectra of the HPP fringes on the right-hand side from the gold crystal edge (Fig. 2(c)) are shown in Fig. 2(d), clearly demonstrating the spatial frequency shift: the peak position of the fitted Lorentzian-like curve (red) is 1.75 μm$^{-1}$, 1.87 μm$^{-1}$, 1.96 μm$^{-1}$, and 2.06 μm$^{-1}$ at $\varphi = 0°$, 45°, 135° and 180°, respectively. The maximum spatial frequency difference between 0° and 180° is 0.29 μm$^{-1}$ which is very close to the analytically estimated value $2\nu_0 \cos\alpha \cos\varphi = 0.26$ μm$^{-1}$ according to Eq. (4).

The extracted HPP spatial frequencies for all measured cases are summarized in Fig. 2(e), clearly demonstrating the blueshift when $\varphi$ increases. The corresponding deviation of the measured spatial frequency is analyzed in Fig. 2(f). The black solid line shows the analytically predicted shift $v_0 \cos\alpha \cos\varphi$, while the measured momentum is shown by the average value at different excitation frequency (normalized by $v_0$). The relative error for the extracted polariton momentum can thus be calculated as an inverse of the effective mode index: $\cos\alpha \cos\varphi / n_{\text{eff}}$.

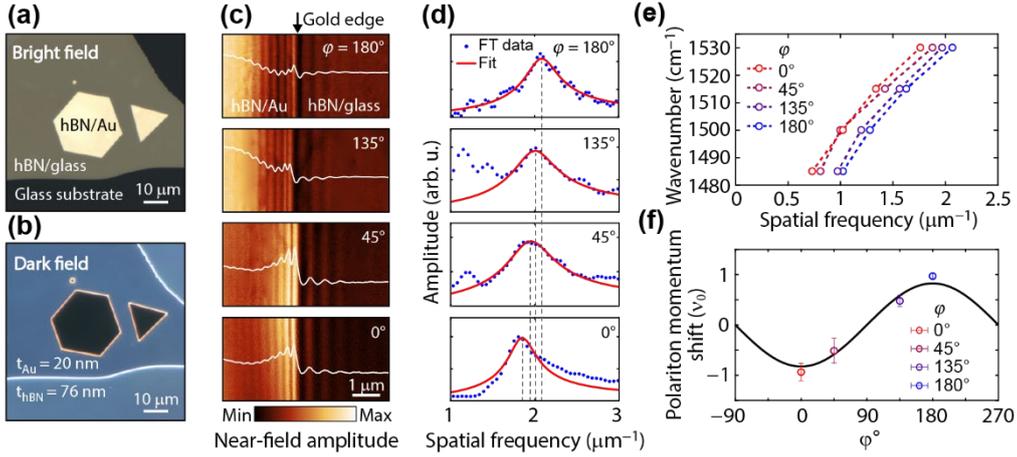

Fig. 2. The relationship between the orientation angle and the spatial frequency shift. (a) Bright field and (b) dark field optical microscope images of a 76 nm-thick hBN flake transferred on top of the monocrystalline gold flake. (c) Near-field amplitude map near the gold crystal edge at different orientation angles $\varphi$. The polariton modes to the left from the edge is the hBN/gold HIP, and the ones to the right is the hBN/glass HPP. (d) Fourier spectra of HPP shown in (c) (blue dots), fitted by the Lorentzian (red line). The spectral peak position of HPP is shifting as the angle changes (gray dashed line). (e) Measured dispersion of HPP at different edge orientation (data points), showing the significant frequency shift for HPP. (f) HPP momentum shift normalized by $v_0$, depending on the edge orientation angle $\varphi$, obtained from the data shown in (e). The data points show the average momentum at given $\varphi$, and the error bar shows the standard deviation. The black solid line is theoretically predicted deviation, $v_0 \cos\varphi \cos\alpha$.

Experimental results show the deviation of the extracted polariton momentum from the actual value by as much as 7.7% even for the HPP with relatively large effective index of 12.5. Therefore, even when $n_{\text{eff}} > 10$, the previously ignored orientation-mediated deviation of the near-field interference must be considered. We also note that the interference fringes due to the HIP appears on the left-hand side from the gold edge (Fig. 2(c)), where the hBN flake is atop the Au crystal. In this region, the surface plasmon on gold (also launched by the gold edge), results in a noticeable fluctuation in the near-field amplitude. The presence of surface plasmons could potentially jeopardize the precise extraction of polariton momentum by interfering with the HPP, making it necessary to exercise special caution in similar experiments.

# IV. POLARITON DAMPING ANALYSIS

We proceed with the methodology for the extraction of the polariton damping from the near-field interference fringes. As discussed earlier, fitting the Fourier spectrum of the near-field interference pattern by Eq. (5) allows to extract not only the spatial frequency $\nu_p$, but also the damping rate of the polariton $\gamma_p$, albeit only for the edge-launched component. With the assumption of $\gamma_p \ll \nu_p$, which holds for polaritons in most van der Waals crystals [40], Eq. (5) provides the maximal Fourier amplitude $|F(\nu_p)| \approx A/2\gamma_p$. By denoting the FWHM as $\Gamma$ and keeping only the lowest-order nonzero terms, we arrive at the identity:

$$\frac{A}{4\gamma_p} \approx \frac{|F(\nu_p)|}{2} = \left|F\left(\nu_p \pm \frac{\Gamma}{2}\right)\right| \approx \frac{A}{2\gamma_p}\sqrt{\frac{1}{1+\left(\frac{\Gamma}{2\gamma_p}\right)^2}}. \qquad (6)$$

To satisfy the equation above, the relationship between FWHM and the damping rate $\gamma_p$ should be

$$\gamma_p = \mathrm{Im}\{\tilde{k}_p\}/2\pi \approx \frac{\sqrt{3}}{6}\Gamma. \qquad (7)$$

It is worth emphasizing that the derived relationship between $\gamma_p$ and $\Gamma$ is largely different from the assumption of $\gamma_p = \Gamma$ that only holds for truly Lorentzian spectra in a form $F_L(s) = 0.5\Gamma/[s^2 + (0.5\Gamma)^2]$.

The relationship described by Eq. (7) can be verified by a full-wave numerical simulation as illustrated in Fig. 3(a). The interference between the TM-polarized incident field $E_i$ and the polaritons it excites at the gold edge produces the fringes of $|E_z|$ (red curves in Fig. 3(b)-(c)), as predicted by Eq. (4). The background signal fluctuation at the HPP side (blue dashed curve in Fig. 3(b)) is caused by the self-interference of incident light upon reflection and is filtered out in s-SNOM experiments (Fig. 2(c)).

Figures 3(d)-(e) show the Fourier spectra of $|E_z|$ with removed background (blue dots), perfectly fitted by the curve given by Eq. (5) (red solid). Fitted damping rate $\gamma_p$ = 0.045 μm$^{-1}$ and 0.071 μm$^{-1}$ for HPP and HIP, respectively, while the FWHM of the Fourier spectra $\Gamma$ = 0.16 μm$^{-1}$ and 0.24 μm$^{-1}$, or $\approx 2\sqrt{3}$ times larger compared to the damping rate as predicted by Eq. (7). Same tendency is observed for the HIP analyzed in Fig. 1: the FWHM values for $\varphi = 0°$ and $180°$ are 0.22 μm$^{-1}$ and 0.24 μm$^{-1}$, respectively, while $\gamma_p$ = 0.062 μm$^{-1}$ and 0.067 μm$^{-1}$, revealing the same $2\sqrt{3}$ difference.

At the same time, Fourier analysis of the interference pattern from the tip-launched polaritons requires more elaborate consideration since it is not rigorously described by the simple expression given by Eq. (5) due to the geometric term. The diverging wavefront of the tip-launched polaritons effectively leads to a faster decay of the field amplitude, resulting in a broader linewidth in the Fourier domain which can be seen in Fig. 1(f).

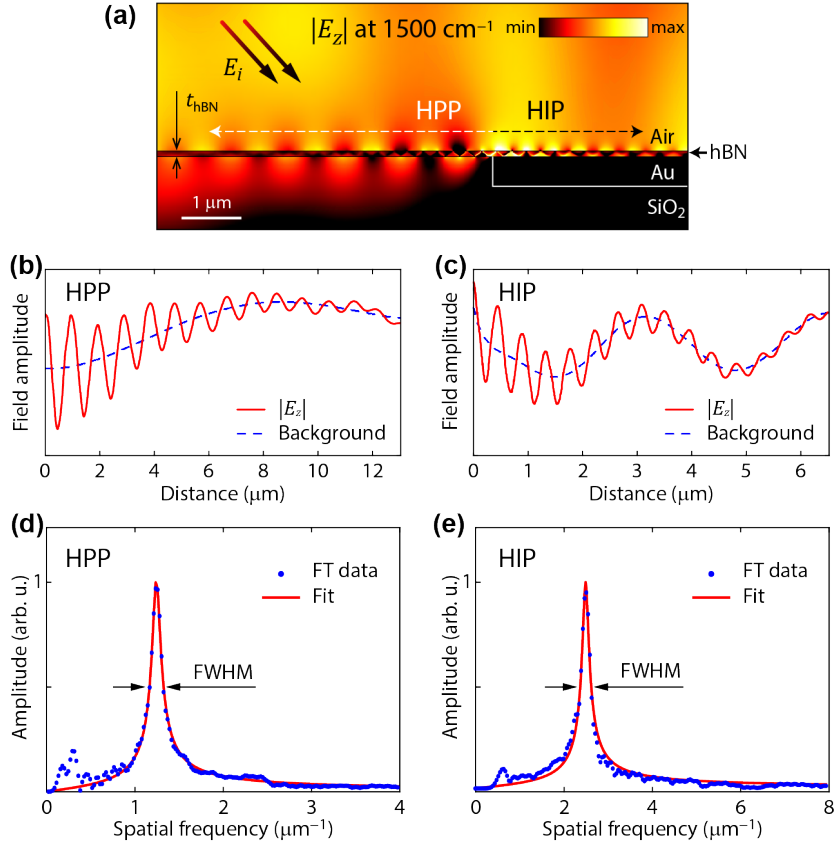

Fig. 3. Comparison of FWHM and fit parameter through simulation results (a) Numerically calculated field interference of the HPP and HIP excited at the gold edge at 1500 cm$^{-1}$; $t_{hBN}$ = 80 nm. The incident plane wave illuminates the shown area at the incident angle $\alpha = 30°$. (b), (c) $|E_z|$ profile (red solid curves) of HPP and HIP, respectively, extracted at 70 nm above the hBN surface (dashed line in (a)). The blue dashed line shows the background of the interference field. (d), (e) Fourier spectra of $|E_z|$ (blue dots) of the HPP and HIP, respectively, fitted by the Lorentzian-like shape given by the Eq. (5) (red curves).

Figure 4(a) illustrates the Fourier spectrum of the near-field fringes above the hBN flake on evaporated gold (inset). To filter out a parasitic near-field signal from the strong scattering at the hBN edge, the interference pattern is analyzed starting from the second maximum (indicated by the white arrow in the inset). The two peaks in the spectrum correspond the edge and tip components, while the tip component is much stronger due to the weaker excitation of the edge component at the given angle $\varphi = 180°$.

Fitting the spectrum of tip component by Eq. (5) yields $\gamma_p$ = 0.228 μm$^{-1}$ and $2\nu_p$ = 4.81 μm$^{-1}$, while the analytical values for HIP mode are $\gamma_{p,HIP}$ = 0.059 μm$^{-1}$ and $2\nu_{p,HIP}$ = 4.76 μm$^{-1}$, indicating the presence of extra damping due to the diverging wavefront and the double travel path.

In order to disentangle the polariton damping from the geometrical decay factor, we analyze the near-field signal in real space. The interference pattern generated purely by the tip-launched polariton (blue dotted line in Fig. 4(b)) is restored from the spectrum in Fig. 4(a) by taking the inverse Fourier transform of the spectrally filtered data (shaded area in Fig. 4(a)). Then, the near-field pattern can be directly fitted by the two functions without (Eq. (8)) and with (Eq. (9)) the geometrical decay factor:

$$f(x) = e^{-2\pi\gamma_p x} \cos\left(\frac{4\pi x}{\lambda_p}\right), \tag{8}$$

$$f(x) = \frac{e^{-2\pi\gamma_p x} \cos\left(\frac{4\pi x}{\lambda_p}\right)}{\sqrt{2(x + \lambda_p/2)}}. \tag{9}$$

Both Eq. (8) and Eq. (9) fit the inversed FT data well, correctly providing the value of $\lambda_p$ = 0.42 μm. However, the fitted damping rates are drastically different. Equation (8) yields $\gamma_p$ = 0.228 μm$^{-1}$, very similar to the fitted parameter of Eq. (5) in momentum space. Equation (9) with the geometric term provides $\gamma_p$ = 0.176 μm$^{-1}$ in a much better agreement with the analytically predicted HIP damping $2\gamma_{p,HIP}$ = 0.118 μm$^{-1}$. Thus, Eq. (9) provides a simple yet potent model to analyze the near-field interference pattern generated by the tip-launched polariton. We emphasize that the spectral filtering of the tip component in the Fourier space is required, followed by the inverse FFT and fitting the real space data.

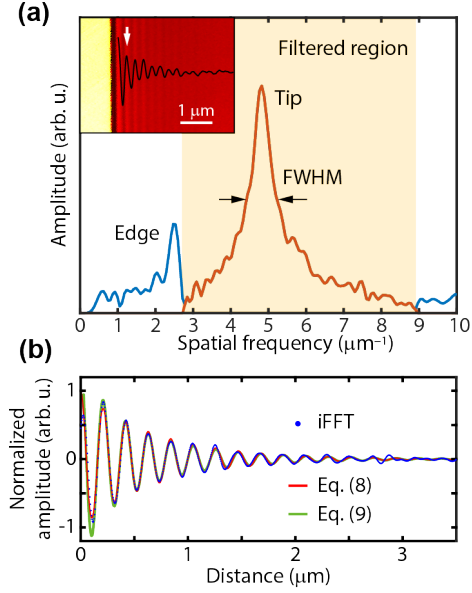

Fig. 4. Damping analysis for the tip component. (a) Fourier spectrum of the near-field amplitude profile from 89 nm-thick hBN on evaporated gold at 1510 cm$^{-1}$ (inset). The signal is truncated at the second maximum (white arrow), and spectrum is filtered (orange shaded region) to isolate the near-field signal from the tip component. (b) Inverse Fourier transform data of the extracted tip component (blue dots) and fitting results with (green) and without (red) the geometric damping term.

## V. CONCLUSION

In summary, we demonstrate rigorous analysis of the near-field interference patterns obtained in the s-SNOM experiments with propagating polaritons, highlighting the important considerations for accurate measurement of the polariton propagation constant. First, we show that the near-field interference pattern can be analyzed in Fourier space, allowing for the independent analysis of several modes typically present in the near-field experiment and accurate extraction of the damping coefficient if the mode is launched by the material edge. Second, we show that the sample orientation may affect the near-field interference pattern and introduce a significant error in the extracted polariton momentum. We experimentally show that this error can be as large as 7.7 % even for modes with effective index exceeding 10. Our results demonstrate the importance of the sample direction in s-SNOM, and the necessity to consider the sample orientation in near-field analysis of the polaritons. Next, we analytically investigate the correlation between the FWHM of the Fourier spectrum and the damping of the propagating polariton, revealing their correlation with a specific ratio $2\sqrt{3}$. Finally, we elucidate the contribution of the geometric term to the damping of the tip component in both real space and Fourier space, emphasizing the crucial importance of the geometric term for fringes analysis.

## ACKNOWLEDGMENTS

This research was supported by the National Research Foundation of Korea (NRF) grants funded by the Ministry of Science and ICT (NRF-2022R1A2C2092095) and the Ministry of Education (NRF-2021R1I1A1A01057510). This work was also supported under the framework of international cooperation program managed by the NRF (2022K2A9A2A23000272, FY2022). The Center for Polariton-driven Light–Matter Interactions (POLIMA) is funded by the Danish National Research Foundation (Project No. DNRF165). F. K. and G. T. acknowledge the support of the Swiss National Science Foundation (Eccellenza Grant #194181). This work was also supported by the BK21 FOUR Program through the NRF funded by Ministry of Education.

## APPENDIX A: SAMPLE PREPARATION AND CHARACTERIZATION

Evaporated gold substrates were made by thermal evaporation method with evaporation velocity of 0.2 Å/s and the sample stage rotated by 1 Hz to reduce surface roughness. Au crystals were grown on borosilicate glass by adding halide ions to a halide-free synthesis recipe [53]. Two glass substrates were cleaned by acetone and ethanol within an ultrasonic bath for 10 min respectively and Tilt the two glass substrates and immerse them into the growth solution. The gap between the two glass substrates was 43 μm, and 50 μL of KBr and 50 μL of KCl solutions (0.5 mM) were added to the growth solution. The solution was stirred and then rested for 1 minute, before being immersing the capped tube in a light-tight water bath at 90°C for 8 to 25 hours. After removing the substrates from the solution, they were rinsed thoroughly with ethanol and DI water to remove loosely attached MFs and remaining reagents. Finally, nitrogen was used blow the samples dry. The hBN bulk material was purchased from a commercial source (2D semiconductors USA), and hBN layers were mechanically exfoliated from the bulk single crystal, and then transferred by the polydimethylsiloxane (PDMS) stamp onto the substrates. Near-field measurements were performed with a commercial s-SNOM (Neaspec GmbH), coupled with the tunable quantum cascade laser (Daylight solutions, MIRcat). The Pt/Ir coated metallic AFM tips (NanoWorld, ARROW-NCPt) were used in tapping mode with a frequency $\Omega$ around 270 kHz and an amplitude of 70 nm. The interferometric signal was demodulated at third harmonic $3\Omega$ to remove the background. The thickness of the hBN flakes was simultaneously measured by the AFM.

# REFERENCES

1. Dai, S., et al., *Tunable phonon polaritons in atomically thin van der Waals crystals of boron nitride.* Science, 2014. **343**(6175): p. 1125-1129.
2. Zheng, Z., et al., *A mid-infrared biaxial hyperbolic van der Waals crystal.* Sci. adv., 2019. **5**(5): p. eaav8690.
3. Álvarez-Pérez, G., et al., *Infrared permittivity of the biaxial van der waals semiconductor α-MoO3 from near-and far-field correlative studies.* Adv. Mater., 2020. **32**(29): p. 1908176.
4. Taboada-Gutiérrez, J., et al., *Broad spectral tuning of ultra-low-loss polaritons in a van der Waals crystal by intercalation.* Nat. mater., 2020. **19**(9): p. 964-968.
5. Chen, S., et al., *Real-space observation of ultraconfined in-plane anisotropic acoustic terahertz plasmon polaritons.* Nat. Mater., 2023: p. 1-7.
6. Chen, J., et al., *Optical nano-imaging of gate-tunable graphene plasmons.* Nature, 2012. **487**(7405): p. 77-81.
7. Fei, Z., et al., *Gate-tuning of graphene plasmons revealed by infrared nano-imaging.* Nature, 2012. **487**(7405): p. 82-85.
8. Hu, F., et al., *Imaging exciton–polariton transport in MoSe2 waveguides.* Nat. Photonics, 2017. **11**(6): p. 356-360.
9. Kim, S., et al., *Functional Mid-Infrared Polaritonics in van der Waals Crystals.* Adv. Opt. Mater., 2020. **8**(5): p. 1901194.
10. Basov, D., M. Fogler, and F. García de Abajo, *Polaritons in van der Waals materials.* Science, 2016. **354**(6309): p. aag1992.
11. Duan, J., et al., *Twisted nano-optics: manipulating light at the nanoscale with twisted phonon polaritonic slabs.* Nano Lett., 2020. **20**(7): p. 5323-5329.
12. Kim, S., et al., *Electrical modulation of high-Q guided-mode resonances using graphene.* Carbon Trends, 2022. **8**: p. 100185.
13. Kim, S., et al., *Ultracompact electro-optic waveguide modulator based on a graphene-covered λ/1000 plasmonic nanogap.* Opt. Express, 2021. **29**(9): p. 13852-13863.
14. Li, P., et al., *Infrared hyperbolic metasurface based on nanostructured van der Waals materials.* Science, 2018. **359**(6378): p. 892-896.
15. Kim, S., et al., *Electronically tunable perfect absorption in graphene.* Nano lett., 2018. **18**(2): p. 971-979.
16. Han, S., et al., *Complete complex amplitude modulation with electronically tunable graphene plasmonic metamolecules.* ACS nano, 2020. **14**(1): p. 1166-1175.
17. Zhang, J., L. Zhang, and W. Xu, *Surface plasmon polaritons: physics and applications.* Journal of Physics D: Appl. Phys., 2012. **45**(11): p. 113001.
18. Sharma, B., et al., *SERS: Materials, applications, and the future.* Mater. today, 2012. **15**(1-2): p. 16-25.
19. Han, H.J., et al., *Synergistic integration of chemo-resistive and SERS sensing for label-free multiplex gas detection.* Adv. Mater., 2021. **33**(44): p. 2105199.

20. Lu, G., et al., *Engineering the spectral and spatial dispersion of thermal emission via polariton–phonon strong coupling.* Nano Lett., 2021. **21**(4): p. 1831-1838.

21. Jaffe, G.R., et al., *The Effect of Dust and Hotspots on the Thermal Stability of Laser Sails.* arXiv preprint arXiv:2303.14165, 2023.

22. Holdman, G.R., et al., *Thermal Runaway of Silicon-Based Laser Sails.* Adv. Opt. Mater., 2022. **10**(19): p. 2102835.

23. Su, X., et al., *Atomic-Scale Confinement and Negative Refraction of Plasmons by Twisted Bilayer Graphene.* Nano Lett., 2022. **22**(22): p. 8975-8982.

24. Jiang, L., et al., *Plasmonic Biosensing with Nano-Engineered Van der Waals Interfaces.* ChemPlusChem, 2022. **87**(11): p. e202200221.

25. Song, K.M., et al., *Microcellular sensing media with ternary transparency states for fast and intuitive identification of unknown liquids.* Sci. Adv., 2021. **7**(38): p. eabg8013.

26. Xu, T., et al., *Plasmonic beam deflector.* Opt. Express, 2008. **16**(7): p. 4753-4759.

27. Siegel, J., et al., *Electrostatic Steering of Thermal Emission with Active Metasurface Control of Delocalized Modes.* arXiv preprint arXiv:2308.07998, 2023.

28. Kim, J.Y., et al., *Full 2π tunable phase modulation using avoided crossing of resonances.* Nat. Commun., 2022. **13**(1): p. 2103.

29. Chen, X., et al., *Modern scattering-type scanning near-field optical microscopy for advanced material research.* Adv. Mater., 2019. **31**(24): p. 1804774.

30. Dai, S., et al., *Subdiffractional focusing and guiding of polaritonic rays in a natural hyperbolic material.* Nat. commun., 2015. **6**(1): p. 6963.

31. Dai, S., et al., *Efficiency of launching highly confined polaritons by infrared light incident on a hyperbolic material.* Nano lett., 2017. **17**(9): p. 5285-5290.

32. Ambrosio, A., et al., *Selective excitation and imaging of ultraslow phonon polaritons in thin hexagonal boron nitride crystals.* Light: Sci. Appl., 2018. **7**(1): p. 27.

33. Giles, A.J., et al., *Ultralow-loss polaritons in isotopically pure boron nitride.* Nat. mater., 2018. **17**(2): p. 134-139.

34. Ni, G., et al., *Fundamental limits to graphene plasmonics.* Nature, 2018. **557**(7706): p. 530-533.

35. Nikitin, A., et al., *Real-space mapping of tailored sheet and edge plasmons in graphene nanoresonators.* Nat. Photonics, 2016. **10**(4): p. 239-243.

36. Bylinkin, A., et al., *Real-space observation of vibrational strong coupling between propagating phonon polaritons and organic molecules.* Nat. Photonics, 2021. **15**(3): p. 197-202.

37. Ma, W., et al., *In-plane anisotropic and ultra-low-loss polaritons in a natural van der Waals crystal.* Nature, 2018. **562**(7728): p. 557-562.

38. Ni, X., et al., *Observation of directional leaky polaritons at anisotropic crystal interfaces.* Nat. Commun., 2023. **14**(1): p. 2845.

39. Menabde, S.G., et al., *Near-field probing of image phonon-polaritons in hexagonal boron nitride on gold crystals.* Sci. Adv., 2022. **8**(28): p. eabn0627.

40. Menabde, S.G., et al., *Low-Loss Anisotropic Image Polaritons in van der Waals Crystal α-MoO3.* Adv.

Opt. Mater., 2022. **10**(21): p. 2201492.

41. Wong, K.P., et al., *Edge-Orientation Dependent Nanoimaging of Mid-Infrared Waveguide Modes in High-Index PtSe2.* Adv. Opt. Mater., 2021. **9**(13): p. 2100294.
42. Kaltenecker, K.J., et al., *Mono-crystalline gold platelets: a high-quality platform for surface plasmon polaritons.* Nanophotonics, 2020. **9**(2): p. 509-522.
43. Mancini, A., et al., *Near-Field Retrieval of the Surface Phonon Polariton Dispersion in Free-Standing Silicon Carbide Thin Films.* ACS Photonics, 2022. **9**(11): p. 3696-3704.
44. Casses, L.N., et al., *Quantitative near-field characterization of surface plasmon polaritons on monocrystalline gold platelets.* Opt. Express, 2022. **30**(7): p. 11181-11191.
45. Caldwell, J.D., et al., *Photonics with hexagonal boron nitride.* Nat. Rev. Mat., 2019. **4**(8): p. 552-567.
46. Kiani, F. and G. Tagliabue, *High Aspect Ratio Au Microflakes via Gap-Assisted Synthesis.* Chem Mater., 2022. **34**(3): p. 1278-1288.
47. Zheng, Z., et al., *Highly confined and tunable hyperbolic phonon polaritons in van der Waals semiconducting transition metal oxides.* Adv. Mater., 2018. **30**(13): p. 1705318.
48. Passler, N.C., et al., *Hyperbolic shear polaritons in low-symmetry crystals.* Nature, 2022. **602**(7898): p. 595-600.
49. Huber, A., et al., *Near-field imaging of mid-infrared surface phonon polariton propagation.* Appl. Phys. Lett., 2005. **87**(8).
50. Lee, I.-H., et al., *Image polaritons in boron nitride for extreme polariton confinement with low losses.* Nat. commun., 2020. **11**(1): p. 3649.
51. Menabde, S.G., et al., *Image polaritons in van der Waals crystals.* Nanophotonics, 2022. **11**(11): p. 2433-2452.
52. Mester, L., A.A. Govyadinov, and R. Hillenbrand, *High-fidelity nano-FTIR spectroscopy by on-pixel normalization of signal harmonics.* Nanophotonics, 2021. **11**(2): p. 377-390.
53. Krauss, E., et al., *Controlled growth of high-aspect-ratio single-crystalline gold platelets.* Cryst. Growth and Des., 2018. **18**(3): p. 1297-1302.